\documentclass[twocolumn,letter]{jpsj2}
\usepackage{amssymb}
\usepackage{epsfig}

\begin{document}
\title{Wilson-like real-space renormalization group and low-energy effective spectrum of the XXZ chain in the critical regime}
\author{Kouichi Okunishi}
%\affiliation{Department of Physics, Niigata University, Igarashi 2, Niigata 950-2181, Japan.}
\inst{Department of Physics, Niigata University, Igarashi 2, Niigata 950-2181, Japan.}
%\address{Department of Physics, Niigata University, Igarashi 2, Niigata 950-2181, Japan.}
\date{\today}
 
%\begin{abstract}
\abst{
We present a novel real-space renormalization group(RG) for the one-dimensional XXZ model in the critical regime, reconsidering the role of the cutoff parameter in Wilson's RG for the Kondo impurity problem. 
We then demonstrate the RG calculation for the XXZ  chain with the free boundary.
Comparing the hierarchical structure of the obtained low-energy spectrum with the Bethe ansatz result, we find that the proper scaling dimension is reproduced as a fixed point of the RG transformation.
}
%\end{abstract}

%\pacs{75.10.Jm, 05.50.+q}
\kword{real-space renormalization group, spin chain, cutoff, Bethe ansatz, scaling dimension}

%%%%%%%%%%%%%%%%%%%%%%%%

\maketitle

%%%%% introduction %%%%%%
  
In the last decade, the density matrix renormalization group(DMRG) has been the most  reliable numerical renormalization group(NRG) for the ground-state of the one-dimensional(1D) quantum many body system.\cite{White}
Before DMRG, however, the conventional NRG scheme such as a block spin transformation often failed for the 1D quantum system.
The reason for this was analyzed carefully by White\cite{dresden}; the effect of the boundary is significant in the 1D system, and his DMRG algorithm overcome it successfully.
On the other hand,  Wilson's NRG procedure for the Kondo impurity problem \cite{wilson,KWW} has been highly successful for solving the various impurity problems.
In Wilson's NRG scheme, the log-discretized energy shells for the s-wave electrons are mapped into an effective tight binding model with a cutoff parameter, which is essentially the 1D quantum many body problem with a boundary.
However, it should be remarked that  Wilson's NRG is efficient for the gapless system, in contrast to the  DMRG which prefers the gapful system.

In this paper, we  focus on the NRG for the gapless 1D quantum spin system; 
Wilson's NRG for the impurity problem is reconsidered in the context of the 1D quantum spin chain.
The essence of the idea is very simple; remove the impurity site and add the interacting spins instead of the free electrons in the impurity problem.
Such an approach without the cutoff parameter was tested by Xiang {\it et al} in the almost the same timing as the DMRG, but it seems more suitable for the gapful spin chain.\cite{xiang}
Also, the infinite-system-size DMRG  still cannot overcome the critical fluctuation in the 1D critical system in the bulk limit;
An essential point in the present approach for the critical system is that we introduce the cutoff parameter controlling the energy scale of the system, which plays an important role in the RG transformation.

In the following, we briefly formulate the Wilson like NRG for the XXZ spin chain, where we emphasize the role of the cutoff parameter.
Numerical calculations are actually performed for the XXZ chain  in the critical region, and the obtained results are analyzed on the basis of the Bethe ansatz/CFT solution.
We then find that the low-energy excitation spectrum of the XXZ spin chain is successfully reproduced as a fixed point of the RG transformation.

In this paper, we deal with the  $S=1/2$ XXZ chain in the critical regime for simplicity, but the formulation for the general case is straightforward.
We write the local Hamiltonian of the XXZ chain as
\begin{equation}
 h_{n,n+1}= S_n^xS_{n+1}^x+ S_n^yS_{n+1}^y+\Delta S_n^zS_{n+1}^z,
\label{xxzh}
\end{equation}
where $\vec{S}$ is the $S=1/2$ spin matrices and $0\le \Delta\le 1$ is assumed.
We then consider the Hamiltonian of $N$ spins with the cutoff $\Lambda$ as
\begin{equation}
H_{N}(\Lambda) =\sum_{n=1}^{N-1}  \Lambda^{N-n-1} h_{n,n+1} , \label{defh}
\end{equation}
for which the free-boundary condition is basically assumed.
If $\Lambda=1$, eq. (\ref{defh}) becomes the uniform XXZ chain with the free-boundary condition.
In Wilson's original NRG for the Kondo problem, there is an impurity at the $n=0$ site, and $h_{n,n+1}$ should be the tight binding free electrons.
In the present case, $n=0$ is the empty site and the interacting spins are adopted as $h_{n,n+1}$.

In order to treat matrices having different sizes in the RG transformation, we use the notation for a $m\times m$ matrix $X$:
\begin{equation}
X^* = I \otimes X,
\end{equation}
where $I$ is the $2\times 2$ identity matrix for the space of the spin added, and thus, $X^*$ becomes a $2m \times 2m$ matrix.  
The recursion relation between the $N$- and $N+1$-spin systems is given by
\begin{equation}
H_{N+1}(\Lambda)= \Lambda H_{N}^{*}(\Lambda) + h_{N,N+1}.
\label{recursion1}
\end{equation}
In this recursion relation, the smallest energy scale is fixed to be in the order of unity.
If $\Lambda=1$, we do not touch the energy scale of the system, leading the simple recursion relation for extending the system size.
For $\Lambda>1$, a new spin having a smaller energy scale is added to the bulk part of the system.
Thus the outer spins(the smaller index of $n$) has the higher energy scale.\cite{index}

Now we convert the basis so as to diagonalize $H_N(\Lambda)$,
\begin{equation}
H_N(\Lambda) U_N= U_N \omega_N, \label{diagh}
\end{equation}
where $\omega$ are the eigenvalues of the Hamiltonian and $U$ are the corresponding eigenvectors.
By maintaining the lower energy states on the basis of (\ref{diagh}), we obtain the recursion relation for the renormalized Hamiltonian:
\begin{eqnarray}
\bar{H}_{N+1}(\Lambda) = \Lambda \omega^*_{N}+ U_{N}^{*\dagger}  h_{N,N+1} U^*_{N},
\label{recursionrg1}
\end{eqnarray}
where $\bar{H}_{N+1}(\Lambda)\equiv U_N^{\dagger*} H^{}_{N+1}(\Lambda)U_N^*$. 
By diagonalizing the Hamiltonian of $N$ sites, we can then increase the system size through eq. (\ref{recursionrg1}) with maintaining lower energy states.
After a sufficient number of iterations starting with the initial Hamiltonian of $N=2$,  the matrices converge to those in the ${\rm N}\to\infty$ limit.
However, we should note that the energy scale of the system is reduced by the cutoff parameter $\Lambda$, which is a crucial point in the RG for the critical system.

\begin{figure}[ht]
\begin{center}
\epsfig{file=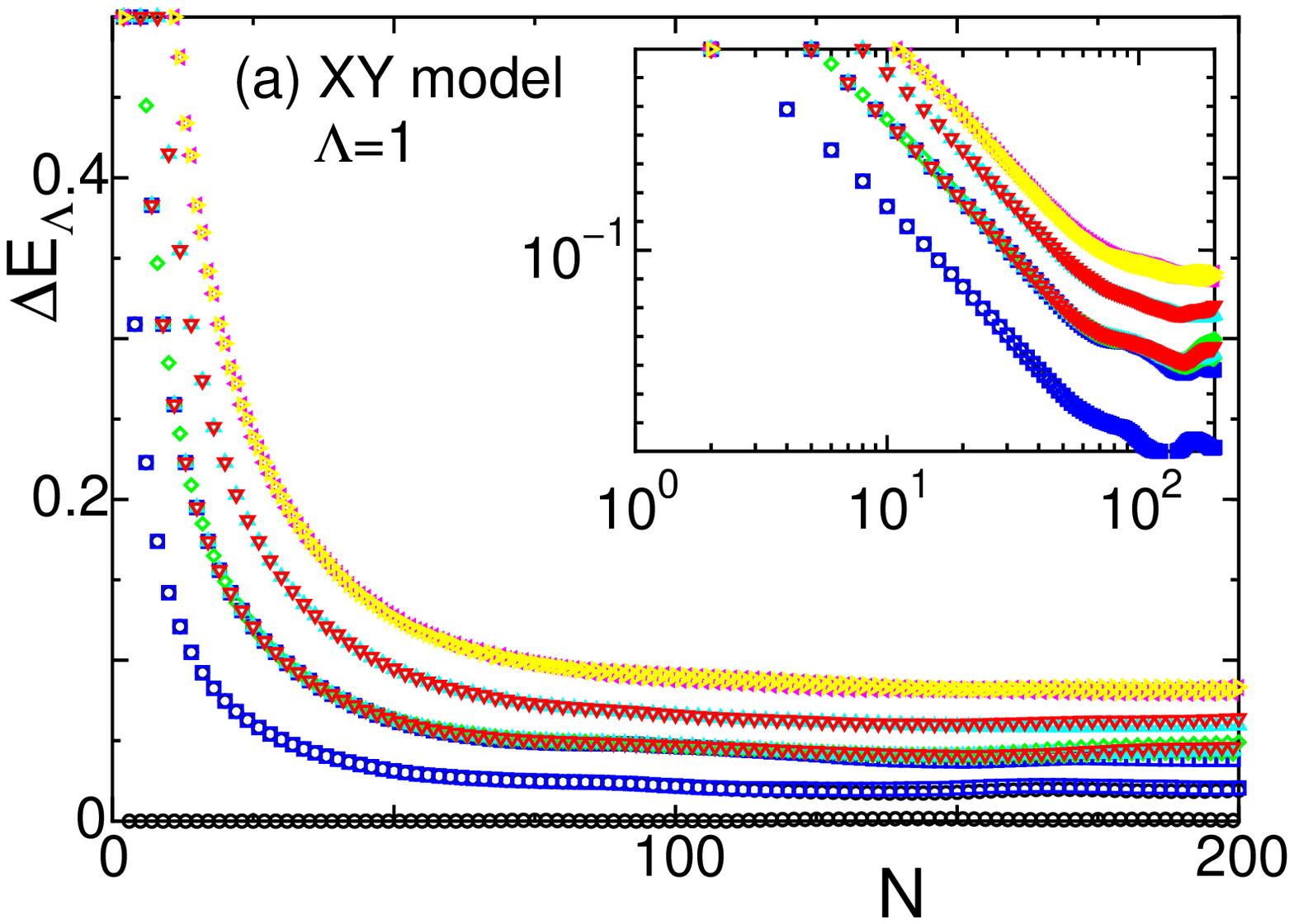,width=7cm}
\epsfig{file=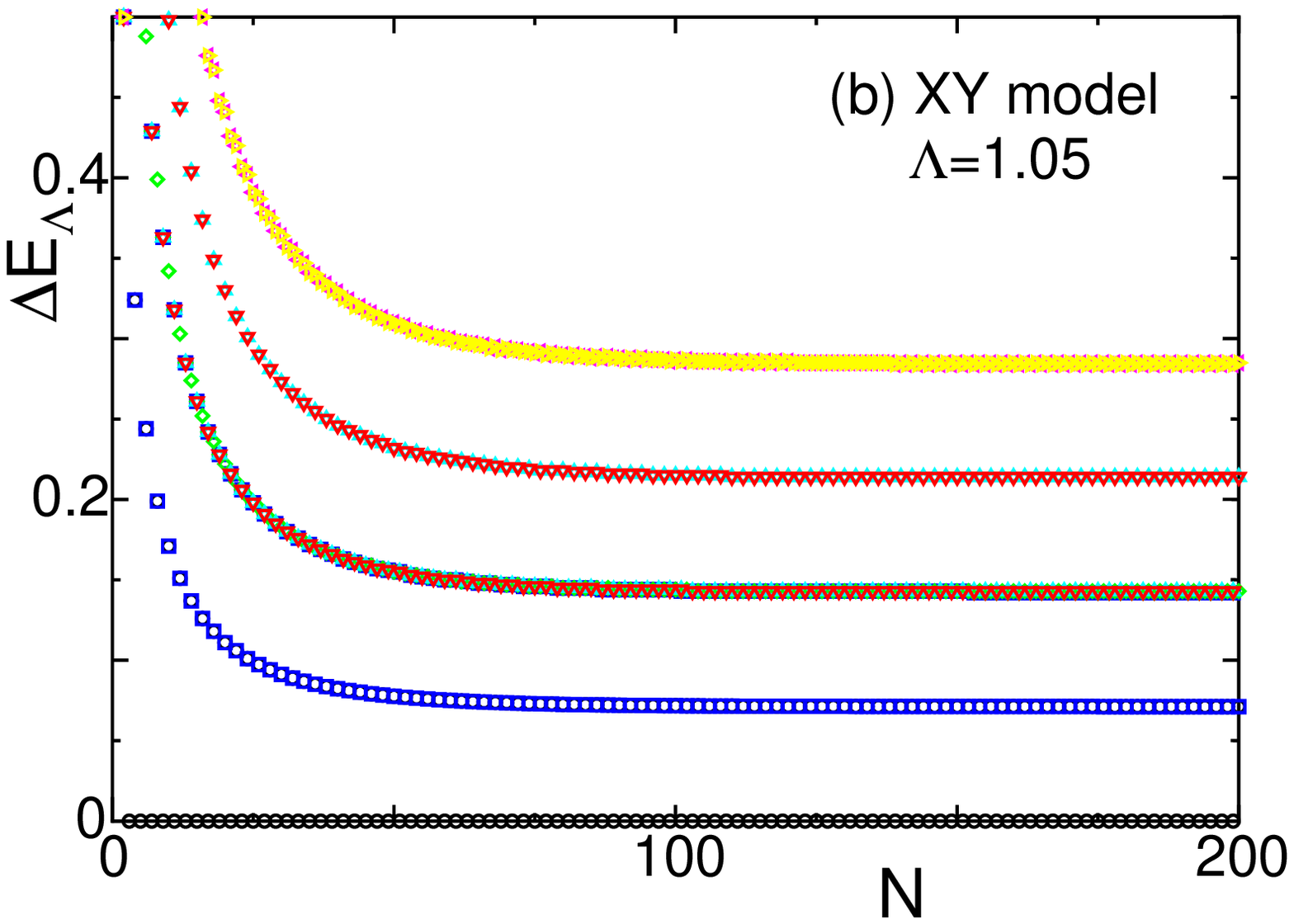,width=7cm}
\end{center}
\caption{Size dependence of the lowest 8  eigenvalues of the Hamiltonian for the XY model: (a) $\Lambda=1$ and (b) $\Lambda=1.05$.
The ground-state corresponds to the horizontal axis.
The inset is a log-log plot of the main panel of (a).}
\label{flow1}
\end{figure}

-{\it XY model $(\Delta=0)$}- Let us demonstrate the NRG for the XXZ chain in the XY-like anisotropic region($0\le\Delta<1$), for which the ground-state has the gapless excitations.
In particular, we first analyze the free fermion case($\Delta=0$).
Figure \ref{flow1} shows $N$-dependence of the lowest eight eigenvalues $\Delta E_\Lambda$ of the XY model with no cutoff($\Lambda=1$) and with cutoff $\Lambda=1.05$. 
In the figures, $\Delta E_\Lambda$ is measured from the ``ground-state'' energy.
We can see the oscillation with respect to $N=$even or odd, for which the total $S^z$ of the system takes an integer or half integer alternately.
Although there are the eight eigenvalues in the figure,  we can see only 5 lines because of the degeneracy. 

As is shown in Fig. 1(a), $\Delta E_\Lambda$ for $\Lambda=1$ shows clear $1/N$ dependence up to $N\sim 50$, which corresponds to  the size scaling of the finite-size energy gap. 
In principle, this implies that the low-energy eigenvalues accumulate to $\Delta E =0$ in the $N \to \infty$ limit.
However, the accuracy for $N>50$ is not maintained because of the truncation of the bases, and thus the lines in Fig. 1(a) show the winding behavior.
We should note that the range of the $1/N$ dependence is limited to up to $N\sim 50$ even with $m=3000$.

In Fig. 1(b), we show the result obtained with the finite cutoff $\Lambda=1.05$, which  converges well with respect to $m(\sim 3000)$. 
In the figure, we can see that the eigenvalues  rapidly approach  constant values.
The regularization due to $\Lambda$ leads to the NRG iteration converging to the fixed point.
Since the energy scale of the system is reduced by $\Lambda$ in each iteration step, this fixed point of the spectrum implies that the system is scale invariant.
More interestingly, the energy levels at the fixed point show very regular behavior;
we can see that the level intervals are constant.
Here, it should be recalled that the smallest energy scale at $N=200$ is of order of $\Lambda^{-200}\sim6\times 10^{-5}$ compared with the largest exchange coupling.

Figure 2 shows the low-energy eigenvalue distribution after sufficient iterations($N=200$).
We can see a clear stairway-like structure having the almost the same level space, which suggests that the fixed point of the free fermion model is properly achieved.
Here we note that this stairway structure is confirmed to be basically the same as that for $\Lambda=1.03$.
In order to analyze the structure, we invoke the finite size spectrum of the XY model, which is obtained by the Jordan-Wigner transformation,\cite{asakawajpa1,asakawa}
\begin{equation}
H_L- E_0\simeq  \frac{\pi v}{L}\left[\sum_{n=0} (n+\frac{1}{2})c_n^\dagger c_n + \sum_{n=0} (n+\frac{1}{2})d^\dagger_n d_n  \right], \label{fssxy}
\end{equation}
where $v$ is the spin wave velocity, $c_n$ and $d_n$ are the fermion operators, and $E_0$ is the ground-state energy including the $1/L$ correction.
Note that the total $S^z$ of the system is represented as  $\hat{M}=\sum_{n=0} c^\dagger_nc_n-d^\dagger_nd_n$.

\begin{figure}[ht]
\begin{center}
\epsfig{file=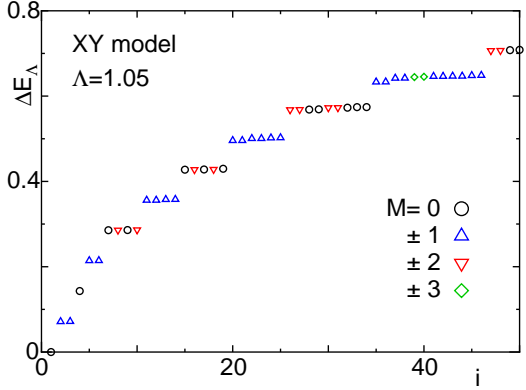,width=7cm}
\end{center}
\caption{Low-energy spectrum of the XY model at the fixed point for $\Lambda=1.05$.}
\label{xyhist}
\end{figure}

The energy scale in the finite-size spectrum is specified by the inverse system size $1/L$, while the energy scale at the fixed point of the NRG is  governed  by the cutoff $\log\Lambda$.
However, the degeneracy of the low-energy excitation in the scaling regime should refer only the scaling dimension.
Applying $H_L$ to the low-energy states labeled by  certain fermion numbers,
we can identify the degenerating structure of eq. (\ref{fssxy}) (See also eq. (9) with $\xi=1$)
\begin{equation}
 {\tt degeneracy\; {\verb/#/} = \{ 1, 2, 1, 2, 4, 4, 5, 6, \cdots \}},
\label{xydegene}
\end{equation}
which is successfully reproduced in Fig. \ref{xyhist}.
In addition, it can be easily verified that the contribution from the sectors labeled by the magnetization quantum number $M$ is also consistent with the exact spectrum.

-{\it XXZ model}- We further analyze the XXZ model, which contains the nontrivial result of the interaction effect.
The exact finite-size spectrum of the XXZ chain with the open-boundary condition in the critical regime was calculated in refs.\cite{asakawajpa2,hamer, asakawa}
In particular, the excitation spectrum for the free-boundary case is given by
\begin{equation}
\Delta E_L(M,N)= \frac{\pi v}{L}Q(M,N),
\label{xxzfss}
\end{equation}
where $v$ is the spin wave velocity and the scaling dimension is defined as
\begin{equation}
 Q(M,N)\equiv \frac{M^2}{2\xi^2} + N .
\label{xxzq}
\end{equation}
The quantum number $M$ is an integer corresponding to the total $S^z$ of the system, and $N$ is a non-negative integer.
In addition, the degeneracy for each $N$ is given by Euler's partition number $\eta(N)$ which is defined by
\begin{equation}
\prod_{n=1}(1-q^n)^{-1}= \sum_{n=0} \eta(n) q^n .
\end{equation}
In eq. (\ref{xxzq}),  the dressed charge $\xi$ can be exactly calculated by the Bethe ansatz for the periodic system,\cite{korepin}
\begin{equation}
 \xi = \left[2(1-\frac{\arccos\Delta}{\pi})\right]^{-1/2},
\label{dcexact}
\end{equation}
which represents the nontrivial consequence of the interaction.
Note that the  dressed charge of the XY model($\Delta=0$) is given by  $\xi=1$, and then eq.(\ref{xxzfss}) can be reduced to eq. (\ref{fssxy}).

\begin{figure}[hb]
\begin{center}
\epsfig{file=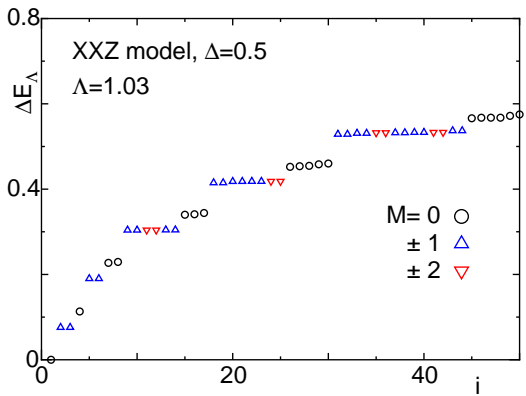,width=7cm}
\end{center}
\caption{The low-energy spectrum of the XXZ model of $\Delta=0.5$ at the fixed point of $\Lambda=1.03$.}
\label{spectrumdelta2}
\end{figure}

Figure \ref{spectrumdelta2} shows the lowest 50 eigenvalues of the XXZ model with $\Delta=1/2$, where the magnetization $M$  of the system is indicated by the distinct symbols.
This spectrum is obtained with the cutoff parameter $\Lambda=1.03$,  after 300 NRG iterations with the number of retained bases $m=3000$. 
Although the degeneracy of the eigenvalues is weakly disturbed by the effect of the finite cutoff, we can clearly recognize the stairway structure.
In Fig. \ref{spectrumdelta2}, an important difference from the free-fermion case is that the level space is not uniform, which is attributed to the dressed charge.

\begin{figure}[ht]
\begin{center}
\epsfig{file=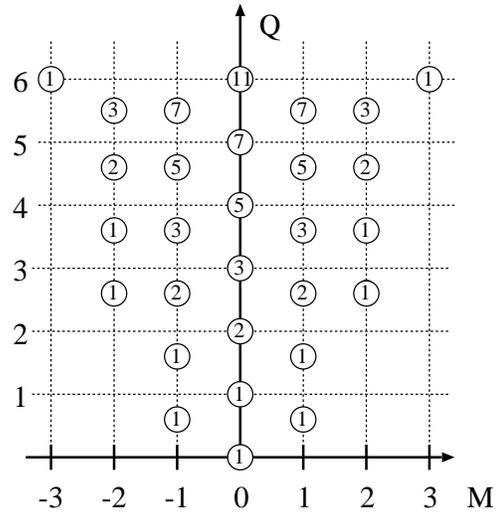,width=6.5cm}
\end{center}
\caption{Tower structure of the scaling dimension $Q(M,N)$ for $\Delta=1/2$.
The numbers in the circle represents the degeneracy numbers $\eta(N)$ for a given set of quantum numbers $(M,N)$.}
\label{tower}
\end{figure}

For the case of $\Delta=1/2$, the exact dressed charge becomes $\xi=\sqrt{3}/2$.
Substituting  $\xi=\sqrt{3}/2$ into eq. (\ref{xxzfss}), we can resolve the stairway structure of the spectrum obtained by NRG;
The tower structure of the weight $Q(M,N)$ for $\Delta=1/2$ is illustrated in Fig. \ref{tower}.
The numbers in the circles indicate the degeneracy corresponding to $\eta(N)$.
Moreover, the lowest energy levels in the $M=\pm 1$ and $\pm 2$ sectors are located at $Q=2/3$ and $8/3$ respectively.
Indeed, the numerical data in Fig. \ref{spectrumdelta2} yields the ratio $\Delta E_\Lambda(0,1)/\Delta E_\Lambda (1,0)=0.68$, which is in good agreement with the exact value $Q=2/3$.
By counting numbers of the degeneracy for each $Q$ level, we obtain 
\begin{equation}
{\tt degeneracy\;{\verb/#/} = \{1,2,1,2,2,6,3,8,5, \cdots \}}.
\end{equation}
These degeneracy numbers are clearly consistent with the stairway structure in Fig. \ref{spectrumdelta2}.
Moreover, in Fig. \ref{spectrumdelta2}, the eigenvalues in  the $M=0$ sector(open circle) clearly show  (quasi)degeneracy corresponding to $\eta(N)$.
The $M=\pm1$ and $\pm 2$ sectors  also exhibit the same hierarchical structure,  starting from $Q=2/3$ and $8/3$ respectively.

\begin{figure}
\begin{center}
\epsfig{file=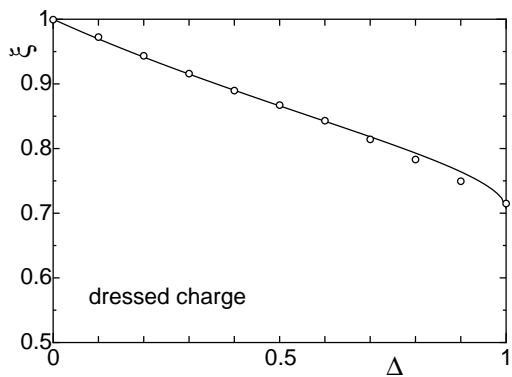,width=6.8cm}
\end{center}
\caption{Dressed charge $\xi$ estimated from the NRG spectrum.
The solid line indicates the exact solution of eq. (\ref{dcexact}). }
\label{figdc}
\end{figure}

We now turn to an application of the NRG.
The ratio of the first to the second excitation energy corresponds to the dressed charge: 
\begin{equation}
\xi = \sqrt{\frac{1}{2}\frac{\Delta E_\Lambda(0,1)}{\Delta E_\Lambda (1,0)}}.
\end{equation}
This relation enables us to estimate the dressed charge from the NRG spectrum.
The  estimated $\xi$ by NRG up to $m=4000$ and $\Lambda=1.02$ is summarized in Fig. \ref{figdc}, where the exact solution is also indicated by the solid line.
The NRG result is in good agreement with the exact solution for $\Delta \le 0.7$, where we have clearly confirmed the stairway structure of the spectrum.
For $0.7<\Delta<1$, however, the open circles slightly deviate from the exact line.
In this region, we can see that the stairway structure in  the NRG spectrum is disturbed significantly within $m=4000$.
For a more accurate estimation of $\xi$, a $\Lambda$ closer to unity and a ``balanced'' increase in $m$ are requested.
However, it should be noted that, at $\Delta=1$, the degeneracy due to the SU(2) multiplets properly emerges in the spectrum, and thus the correct value of $\xi=1/\sqrt{2} (\simeq0.7071)$ is obtained. 
This is because the SU(2) symmetry is well maintained in the NRG iterations, although the cutoff $\Lambda$ is introduced.

To summarize,  we have formulated the Wilson-like real-space RG  for the 1D XXZ model by introducing the cutoff parameter $\Lambda$.
We have demonstrated the NRG calculation for the XXZ chain for $0\le \Delta \le 1$.
Then it was confirmed that the hierarchical degenerating structure of the low-energy spectrum is consistent with  the exact scaling dimension which is calculated by the Bethe ansatz for the free boundary.
Moreover, the estimated dressed charge is in good agreement with the exact value.
These imply that Tomonaga-Luttinger liquid universality can be reproduced as the fixed point of the real-space RG transformation.
In Wilson's original NRG for the impurity problem, the cutoff $\Lambda$ comes from the discretization of the Fermi sea in the vicinity of the Fermi surface.
On the other hand, the cutoff in the present NRG for the XXZ model was directly introduced by hand.
In both cases, the cutoff parameter successfully controls the energy scale of the system.
Nevertheless, we should note that the cutoff also induces an undesirable perturbation in the degenerating structure, in addition to the desired regularization of the gapless spectrum.
In order to extract the correct scaling dimension, this perturbation must be irrelevant compared with the level spaces of the desired  hierarchical structure.
In principle, we can take a sufficiently small $\Lambda(\to 1)$, but the systematical $\Lambda$ dependence is still unknown.
In this sense, further investigation of the $\Lambda$ dependence is clearly needed.

As can be expected, it is also straightforward to apply the present NRG to the gapful spin  chain, and we have actually examined the XXZ chain with $\Delta>1$.
However, the obtained spectrum does not have a regular distribution, and thus we can not resolve any algebraic structure at the present stage.
This may be because the regularization by $\Lambda$  competes with the intrinsic energy gap in the original XXZ chain.
Whether we can extract some information on the gapful system using the present NRG is an interesting future problem.

Finally, we make a comment on the infinite-system-size method of the DMRG.
Of course, the most significant difference of the DMRG from Wilson's NRG is that the RG transformation is constructed through the diagonalization of the reduced density matrix.
However,  another important aspect particularly for the critical system is that the DMRG does not seem to include any scale transformation in its algorithm(the system size increases linearly in DMRG).
In other words, there is no explicit cutoff parameter that controls the energy scale of the system.
In this sense, we may say that the fixed point of the infinite-system-size DMRG is the self-consistent solution of a certain matrix product type wavefunction, apart from the usefulness of the DMRG.
In order to resolve the critical behavior in the DMRG,  it may be a hopeful way that the direct RG analysis of the reduced density matrix or the corner transfer matrix\cite{Baxter}, which show the similar degenerating structure in their eigenvalue spectrum\cite{Baxter,Kaulke,oha,CH}.

%\acknowledgments

This work is supported by Grants-in-Aid for Scientific Research (No.18740230, No.17340100) from MEXT. 
It is also supported by a Grant-in-Aid for Scientific Research in Priority Area ``High Field Spin Science in 100T''.
Part of the numerical computation was performed on SX8 at Yukawa Institute, Kyoto University.

%%%%%%%%%%%%%%%%%%%%%%%%%%%%%%%%%%


\begin{thebibliography}{99}
\bibitem{index} The order of the spin index is different from the previous paper\cite{CH}. We assign $N+1$ to the added spin.
\end{thebibliography}
\end{document}